\documentstyle[graphicx]{article}

\begin{document}
\title{Dirac's principle in multimode interference of independent sources}
\author{P Sancho \\ GPV de Valladolid \\ Centro Zonal en Castilla y Le\'on  \\
Ori\'on 1, 47014, Valladolid, Spain}
\maketitle
\begin{abstract}
The extended Dirac's principle describes the interference between
different particles as an effect of the multiparticle system with
itself. In this paper we present a novel example, based on the
detection of particles emitted in multimode states by independent
sources, which illustrates in a simple way the necessity of
extending the original Dirac's criterion.
\end{abstract}
\newpage
\section{Introduction}
Interference effects are ubiquitous in quantum mechanics. At the
beginning of modern quantum theory Dirac enunciated a criterion
for the existence of interference effects between photons \cite{Dir}:
\begin{quotation}
Each photon interferes only with itself. Interference between two
different photons never occurs.
\end{quotation}
Originally Dirac formulated the principle only for photons, but soon
it was extended to the case of massive particles.

Later, following the analysis of some new developments in quantum
theory it was realized that the criterion should be interpreted in a
more subtle way. These developments refer to situations where the
wave function describing a multiparticle system cannot be separated
into the product of the wave functions of the particles composing
the complete system. There are two well-known scenarios where these
states appear. One is by preparing the particles in an initially
entangled state. In the language of wave functions this property
translates into the impossibility of separating the wave function of
the complete system. The other scenario is related to the
spin-statistics connection. The multiparticle wave function of
bosons or fermions must be symmetrized or antisymmetrized resulting
in non separable wave functions. In both scenarios it is impossible
to speak about the properties of any of the particles as an
independent entity. One is tempted to interpret the interference
effects associated with these systems as interferences between
different particles. However, as the properties of the individual
particles are only defined within a larger entity, the complete
multiparticle system (the only entity that is defined from the
quantum point of view), we really observe interferences of the
complete system. According to Silverman, this extended Dirac's
criterion can be expressed as \cite{Sil}:
\begin{quote}
A system interferes only with itself.
\end{quote}
The extended formulation reflects the fundamental property of
entangled systems of loosing the individuality of the particles
within the multiparticle system, the only {\em single entity} from
a quantum point of view.

Recently, it has been discussed in the literature other experimental
scenario where the extended interpretation must be used, in spite of
the fact that the situation refers to non-entangled photons emitted
by independent sources \cite{Shi}. Two sources emit independently photons,
which are detected at two different positions. When the composed
paths (the paths of the two photons) between the two sources and the
two detectors are indistinguishable we must add the probability
amplitudes obtaining interferences in the joint detection
probability. Once more we do not face interferences between both
photons, since the probability amplitudes that we add are those of
composed paths. The composed paths are not properties of the
individual photons, but a property of the two-photon system.

We present in this paper a new example of multiparticle system for
which it is necessary the extended interpretation. It is based on a
recently proposed arrangement \cite{San}, in which two sources
independently emit particles in non-entangled multimode states. When
the particles have common modes the probability of detecting only
one of the two particles shows interference effects that must be
associated with the complete multiparticle system.

\section{Interferences in multimode states}

We briefly describe the ideal arrangement considered in this paper,
which has been presented in detail in Ref. 4 (see Fig 1). Two
different sources independently emit indistinguishable particles. By
the matter of simplicity we restrict our considerations to bosons
(the case of fermions can be developed along very similar lines, but
needs from a more elaborated and lengthy discussion of some
technical aspects \cite{San}). In the region of overlapping of the two
beams we place at a fixed position a detector. We study the
detections that occur at a given time $t$. We concentrate on the
cases where only one of the two particles is detected, disregarding
the events when the two particles are detected simultaneously. We
assume, by simplicity, that the detector can distinguish between
one- and two-particle detection events (see Ref. 4 for a realistic
arrangement with this property).

\begin{figure}[p]
\center
\includegraphics[width=12cm, height=18cm]{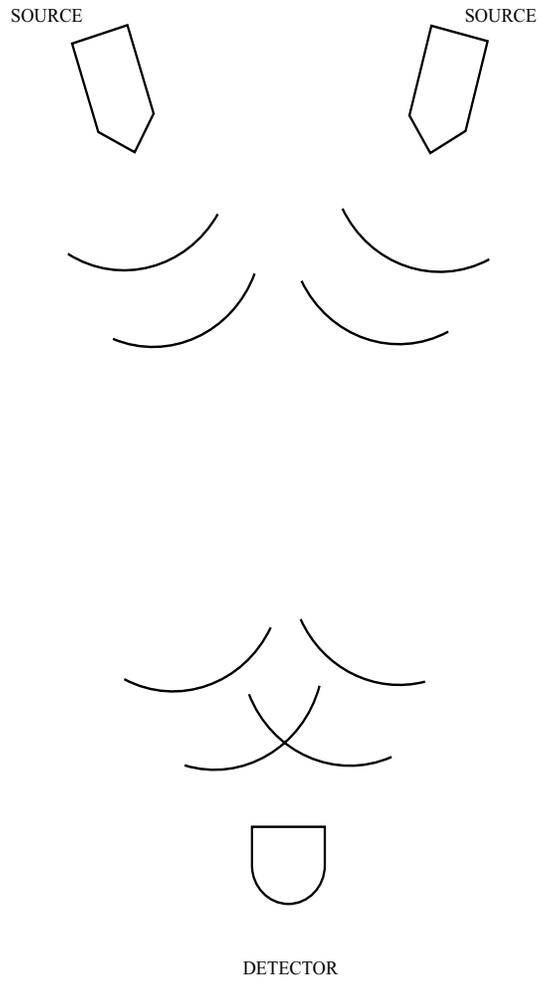}
\caption{ Schematic representation of the experiment. The two
sources emit particles in multimode states characterized by the
complex mode distributions $\mu$ and $\eta$. Both beams come
together and mix at the position of the detector.}
\end{figure}

The particles are emitted by the sources in the state
\begin{equation}
|I>=\int d^3 {\bf p} \int d^3 {\bf q} \eta ({\bf p}) \mu ({\bf q})
\hat{a} ^+ ({\bf p}) \hat{a} ^+ ({\bf q}) |0> , \label{eq:uno}
\end{equation}
where ${\bf p}$ and ${\bf q}$ represent the momenta of the
particles, $\eta $ and $\mu $ are the complex distributions of
momenta, $\hat{a} ^+({\bf p})$ is the creation operator of a
particle with momentum ${\bf p}$ and $|0>$ refers to the vacuum
state in Fock's space. Moreover, by simplicity, we have assumed both
particles to be in the same state of spin. The extension to
particles in different states is straightforward.

The form of Eq. (\ref{eq:uno}) follows directly from the independent
nature of the two sources. When the momenta distributions are
non-trivial, i. e., when they are different from zero for several
values of the momentum we have a multimode distribution. We assume
the modes to be plane waves, i. e., the mode of momentum ${\bf p}$
is $(2 \pi \hbar )^{-3/2} \exp{(i {\bf p}.{\bf r}/\hbar )}$. We
assume both mode distributions to be normalized to unity:
\begin{equation}
\int |\eta ({\bf p})|^2 d^3 {\bf p} =1= \int |\mu ({\bf p})|^2 d^3 {\bf p}.
\label{eq:dos}
\end{equation}
Using the commutation relations (both particles are identical
bosons), $\hat{a} ({\bf p}) \hat{a}^+ ({\bf q}) - \hat{a}^+ ({\bf
q}) \hat{a}({\bf p})=\delta ^3 ({\bf p} -{\bf q})$, and the
property of the annihilation operators $\hat{a} ({\bf p})|0> =0$,
we have
\begin{equation}
<I|I> =1+ \int d^3 {\bf p} \int d^3 {\bf q} \eta ^* ({\bf p}) \mu
^* ({\bf q}) \eta ({\bf q}) \mu ({\bf p}). \label{eq:tres}
\end{equation}
To obtain this relation we have also used the normalization
conditions (\ref{eq:dos}) and the usual normalization of the
vacuum, $<0|0>=1$.

The probability of detecting only one of the two particles at point
${\bf r}$ and time $t$ is given by \cite{Gal}:
\begin{equation}
P({\bf r},t)=\frac{<I| \hat{\psi }^+({\bf r},t) \hat{\psi } ({\bf
r},t)|I>}{<I|I>} , \label{eq:cuat}
\end{equation}
where $\hat{\psi } ({\bf r},t)$ is the Schr\"{o}dinger field
operator, which in the plane wave representation for the modes is
given by
\begin{equation}
\hat{\psi }({\bf r},t) = \frac{1}{(2 \pi \hbar )^{3/2}} \int d^3
{\bf Q } \exp{(i {\bf Q}. {\bf r}/\hbar )} \exp{(-i E({\bf
Q})t/\hbar )}\hat{a} ({\bf Q}),
\end{equation}
where $E({\bf Q})= {\bf Q}^2/2m$ is the energy of the mode ${\bf
Q}$.

Equation (\ref{eq:cuat}) can be evaluated in the same way done for
the denominator (Eq. (\ref{eq:tres})). Interchanging the parameters
${\bf Q}$, ${\bf q}$ and ${\bf p}$ where necessary the result is

\begin{eqnarray}
P({\bf r}, t)= \frac{1}{<I|I>} \frac{1}{(2 \pi \hbar )^3} \int d^3
{\bf p} \int d^3 {\bf q} \eta ^* (\bf p)
\mu ^* (\bf q) \times \nonumber \\
( \exp{(-i({\bf p}.{\bf r}-E({\bf p})t)/\hbar )} \eta ({\bf q}) \int
d^3 {\bf Q} \exp{(i({\bf Q}.{\bf r}-E({\bf Q})t)/\hbar )}
\mu ({\bf Q}) + \nonumber \\
\exp{(-i({\bf q}.{\bf r}-E({\bf q})t)/\hbar )} \eta ({\bf p}) \int
d^3 {\bf Q} \exp{(i({\bf Q}.{\bf r}-E({\bf Q})t))/\hbar )}
\mu ({\bf Q}) + \nonumber \\
\exp{(-i({\bf p}.{\bf r}-E({\bf p})t)/\hbar )} \mu ({\bf q}) \int
d^3 {\bf Q} \exp{(i{\bf Q}.{\bf r}-E({\bf Q})t)/\hbar )}
\eta ({\bf Q})  + \nonumber \\
\exp{(-i({\bf q}.{\bf r}-E({\bf q})t)/\hbar )} \mu ({\bf p}) \int
d^3 {\bf Q} \exp{(i({\bf Q}.{\bf r}-E({\bf Q})t)/\hbar )}
 \eta ({\bf Q})).
\label{eq:siet}
\end{eqnarray}
We rearrange these expressions in a more tractable way. We introduce
the following notation:
\begin{equation}
\alpha _{\mu \eta} =\frac{ \int d^3 {\bf Q} \mu ^* ({\bf Q}) \eta
({\bf Q})}{<I|I>}.
\end{equation}
From the definition and Eq. (\ref{eq:dos}) they follow easily
relations $\alpha _{\eta \eta } = \alpha _{\mu \mu} = <I|I>^{-1}$
and $\alpha _{\mu \eta }^* = \alpha _{\eta \mu}$. These coefficients
correspond to the integrals in Eq. (\ref{eq:siet}) that do not
contain any dependence on ${\bf r}$ or $t$.

On the other hand, we introduce the time and position dependent
functions:
\begin{eqnarray}
P_{\eta \mu } ({\bf r},t) = \\
\frac{1}{(2 \pi \hbar )^3} \int d^3 {\bf p} \int d^3 {\bf q}
\exp{(i(({\bf q}-{\bf p}).{\bf r}-(E({\bf q})-E({\bf p}))t))/\hbar
)} \eta ^* ({\bf p}) \mu ({\bf q}). \nonumber
\end{eqnarray}
Note that $P_{\eta \mu}^* =P _{\mu \eta }$.

Combining all these expressions, we obtain for the detection probability:
\begin{eqnarray}
P({\bf r},t) =\alpha _{\mu \eta } P _{\eta \mu} ({\bf r},t) + \alpha  _{\eta \eta } P _{\mu \mu} ({\bf r},t) + \nonumber \\
\alpha _{\mu \mu } P _{\eta \eta} ({\bf r},t) + \alpha _{\eta \mu } P _{\mu \eta} ({\bf r},t).
\end{eqnarray}
Using the properties of the $\alpha 's$ and $P's$ above remarked we have finally the equation
\begin{equation}
P({\bf r},t) = \alpha _{\eta \eta } P _{\mu \mu} ({\bf r},t) + \alpha
_{\mu \mu } P _{\eta \eta} ({\bf r},t)+ 2Re(\alpha _{\mu \eta } P
_{\eta \mu} ({\bf r},t)), \label{eq:once}
\end{equation}
where $Re(\xi )$ denotes the real part of the complex expression
$\xi $.

Equation (\ref{eq:once}) shows the existence of an interference
phenomenon. The detection probability is composed of three terms.
$P_{\mu \mu }({\bf r},t)$ and $P_{\eta \eta } ({\bf r},t)$ represent
the contributions, up to normalization factors, that one would
obtain if only one of the two sources would emit particles. On the
other hand, the third term $2Re(\alpha _{\mu \eta } P _{\eta \mu
}({\bf r},t))$ has the typical form of an interference term. It
introduces a non trivial deviation with respect to the probability
one would obtain if the detection events would be independent for
both sources: the probability is not the sum of the probabilities
corresponding to each source.

We emphasize that the interference phenomenon is observed at any
fixed point ${\bf r}$. We are not dealing with spatially extended
interferences fringes, for whose observation we would need to move
the detector. The interference effects manifest at every point by
the deviation of the detection probability from that corresponding
to the two sources emitting independently. The form of the third
term depends on the product of the two mode distributions. The
interference phenomenon is clearly dependent on the existence of
common modes: when there are no common modes the product $\eta ^*
({\bf p}) \mu ({\bf p})$ is zero, becoming null the function $P
_{\eta \mu }$ and the interference term.

\section{Discussion}

The interference effects found in the previous section are
explained in terms of the existence of common modes. The detector
cannot distinguish if the common modes belong to one or other of
the particles. In presence of alternatives
that cannot be distinguished quantum mechanics
leads to interference effects.

This is a novel manifestation of the extended Dirac's principle. The
interference effects cannot be associated with one or other of the
particles or with the interaction between them, but with the
existence of common modes, which is not a property of the particles
composing the system, but of the complete system. We face a
self-interference effect of the complete system.

We note that the example proposed in this paper cannot be
interpreted as a new manifestation of the spin-statistics
connection. Although this connection has been used in the
commutation relations it is clear that, in absence of common modes,
the connection does not produce interference effects. We also remark
that in our example the state is a non-entangled one, as in Ref. 3.
The fundamental difference between both arrangements is that in that
case the interference is associated with indistinguishable
two-photon paths, whereas in our example the cause of the
interferences is the existence of common modes.

{\bf Acknowledgments}

This work has been partially supported by the DGICyT of the Spanish
Ministry of Education and Science under Contract n. REN2003-00185.

\end{document}